\documentclass[a4paper,fleqn,12pt]{article}

\newtheorem{theorem}{Theorem}[section]
\newtheorem{definition}{Definition}[section]

\newtheorem{lemma}[theorem]{Lemma}

\newtheorem{hypotheses}{Hypotheses}[section]

\usepackage[T1]{fontenc}
\usepackage[english]{babel}
\usepackage{amsfonts}
\usepackage{amsmath}
\usepackage[left=20mm, right=20mm, top=25mm, bottom=25mm]{geometry}
\usepackage[pdftex]{hyperref}
\usepackage{cite}

\title{Regular Solutions to the Coagulation Equations with Singular Kernels}
\author{Carlos Cueto Camejo\thanks{Corresponding author. Fax: +49 391 6718073,\newline  E-mail address: cuetacam@st.ovgu.de (C. Cueto Camejo).}, Robin Gr\"opler, Gerald Warnecke \\ \small{\emph{Institute for Analysis and Numerics, Otto-von-Guericke University Magdeburg,}} \\ \small{\emph{Universit\"atsplatz 2, D-39106 Magdeburg, Germany}}}

\begin{document}
\maketitle

\begin{abstract}
In this article we prove the existence of solutions to the coagulation equation with 
singular kernels. We use weighted $L^1$-spaces to deal with the singularities in order 
to obtain regular solutions. The Smoluchowski kernel is covered by our proof. The  
weak $L^1$ compactness methods are applied to suitably chosen approximating equations 
as a base of our proof. A more restrictive uniqueness result is also mentioned.
\end{abstract}


\section{Introduction}
Certain problems in the physical sciences are governed by the coagulation equation, 
which describes the kinetics of particle growth where particles can coagulate to form 
larger particles via binary interaction. The coagulation equation was formulated by 
Smoluchowski (1917)\cite{Smoluchowski} and by M\"uller (1928)\cite{Muller} in a discrete 
and an integral form respectively. Examples of this process can be found e.g.\ in 
astrophysics \cite{Dominik}, in chemical and process engineering \cite{Ramkrisha}, and 
aerosol science \cite{Seinfeld}.

Let the non-negative variables $x$ and $t$ represent the size of some particles and 
time respectively. By $u(x,t)$ we denote the number density of particles with size $x$ 
at time $t$. The rate at which particles of size $x$ coalesce with particles of 
size $y$ is represented by the coagulation kernel $K(x,y)$.

The general coagulation equation is now given by
\begin{eqnarray}
\frac{\partial u(x,t)}{\partial t}=\frac{1}{2}\int\limits_0^x K(x-y,y)u(x-y,t)u(y,t)\,dy-\int\limits_0^\infty K(x,y)u(x,t)u(y,t)\,dy.
\label{problema}
\end{eqnarray}

The equation (\ref{problema}) is considered for some given initial 
data $u_0(x)\geq 0$, i.e.\ we consider the initial condition

\begin{eqnarray}
u(x,0)=u_0(x)\geq 0\quad \mbox{a.e.}
\label{condinicial}
\end{eqnarray}

There are many previous results related to the existence and uniqueness of solutions 
to the different forms of the coagulation equation for non-singular kernels, see 
e.g.\ \cite{Fournier}, \cite{McLeod}, \cite{Menon}. But to our knowledge there are 
few works on Smoluchowski's coagulation equation with singular kernels, 
see e.g.\ \cite{ScobedoMischler}, \cite{FournierLaurencot}, \cite{Norris}. Fournier 
and Lauren\c{c}ot \cite{FournierLaurencot} proved the existence of self similar 
solutions to the Smoluchowski coagulation equation with homogeneous kernels while 
Escobedo and Mischler \cite{ScobedoMischler} gave some regularity and size properties 
of the self similar profiles. These special solutions are not a topic of this paper. 
Norris \cite{Norris} proved the existence of weak solutions that are local in time to 
the Smoluchowski equation when the kernel is estimated by the product of sublinear 
functions, i.e.
\begin{eqnarray*}
K(x,y)\leq\varphi(x)\varphi(y)\quad\mbox{with}\quad \varphi:E\rightarrow[0,\infty[, \; \varphi(\lambda x)\leq\lambda\varphi(x)\quad\mbox{for all}\quad x\in E,\lambda\geq 1.
\end{eqnarray*}
In this paper we present a proof of an existence theorem of solutions to the Smoluchowski 
coagulation equation (\ref{problema}) for the following class of singular kernels
\begin{eqnarray} 
K(x,y)\leq k(1+x+y)^\lambda(xy)^{-\sigma}, \qquad\lambda-\sigma\in[0,1[, \;\sigma\in[0,1/2].
\label{asterisco1}
\end{eqnarray}
A key ingredient for our existence theorem is the use of specific weighted $L^1$-spaces. 
Weighted $L^1$-spaces have been used to show existence of solutions to the coagulation-fragmentation
equation, see e.g.\ \cite{Ankik}, \cite{Stewart}.
For our result we introduce the weighted space $L^1\left([0,\infty[;x^{-1}+x\;dx\right)$ for the 
initial data. It is important to point out that our result is also valid for inital data in the weighted space
$L^1\left([0,\infty[;x^{-2\sigma}+x\;dx\right)$ with $\sigma$ as above which in the case of nonsingular 
kernels, i.e.\ $\sigma=0$, becomes $L^1\left([0,\infty[;x\;dx\right)$.

Our existence result is stronger than the result of Norris \cite{Norris} in the following sense. The solutions 
he obtained are weak measure solutions on space and time while our solutions are \emph{regular} 
solutions that lie in the space $C_B^1\left([0,\infty[,L^1\big(]0,\infty[\big)\right)$. 
But note that price we have to pay is that we are more restrictive than Norris on the initial data. Also 
note that our regularity result is obtained in $L^1$ and not in the weighted space. 
So our existence result is less general concerning the initial data but more precise concerning the resulting solutions. 

We call a solution conservative if the total mass of the system remain constant 
throughout time, i.e.
\begin{eqnarray*}
\int\limits_0^\infty xu(x,t)dx=\int\limits_0^\infty xu(x,0)dx\quad\mbox{for all $t\geq 0$.}
\end{eqnarray*}

We would also like to point out that the solutions obtained in Norris \cite{Norris} are 
conservative if $\varphi(x)\geq \varepsilon x$ for all $x$ and some $\varepsilon>0$ and 
\begin{eqnarray}
\int\limits_0^\infty\varphi^2(x)\mu_0(dx)<\infty.
\label{Norrissecondorder}
\end{eqnarray}
These two conditions together mean that he needs at least to bound the second moment 
to have conservative solutions. It can be shown that we just need the $\zeta$-moment 
bound, with $\zeta=1+\lambda-\sigma$ which is a lower moment. 

Our result is obtained in the space $C^1_B\left([0,\infty[, L^1\big(]0,\infty[\big)\right)$ for 
kernels with singularities on the axes, covering in this way the important Smoluchowski 
coagulation kernel 
\begin{eqnarray}
K(x,y)=(x^{1/3}+y^{1/3})(x^{-1/3}+y^{-1/3})
\label{Ksmoluchowski}
\end{eqnarray} 
for Brownian motion, see Smoluchowski ~\cite{Smoluchowski}. This kernel is one of the 
few kernels used in applications that is derived from fundamental physics and not just 
by ad hoc modeling. The equi-partition of kinetic energy (EKE) kernel 
\begin{eqnarray}
K(x,y)=(x^{1/3}+y^{1/3})^2\sqrt{\frac{1}{x}+\frac{1}{y}},
\label{Keke}
\end{eqnarray}
and the granulation kernel
\begin{eqnarray*}
 K(x,y)=\frac{(x+y)^a}{(xy)^b},
\end{eqnarray*}
see Kapur \cite{Kapur}, are also covered by our analysis. These kernels \ref{Ksmoluchowski}  
and \ref{Keke} were not included in the results of Fournier and Lauren\c{c}ot \cite{Fournier}, 
as the authors point out. 

Our approach is based on the well known method by Stewart \cite{Stewart} for 
non-singular kernels. However, it turned out that our modification using weighted 
$L^1$-spaces was not always straight forward. Stewart in his method defined a 
sequence of truncated problems. He proved the existence and uniqueness of solutions 
to them. Using weak compactness theory, he proved that this sequence of solutions 
converges to a certain function. Then it is shown that the limiting function solves 
the original problem. In our approach we redefine Stewart's truncated problem in 
order to eliminate the singularities of the kernels. Using the contraction mapping 
principle we prove that our truncated problems have a unique solution.  We construct 
a singular sequence around the origin to deal with the singularities of the kernels 
and prove that this sequence and the sequence of solutions to the truncated problems 
are weakly relatively compact and equicontinuous in time by using the Dunford-Pettis 
and Arzela-Ascoli Theorem repectively. These properties of the sequence are later 
used to prove that the sequence of solutions to the truncated problem converges to a 
solution of our original problem. In that way we obtain the existence of solutions to 
the coagulation equation with singular kernels. The uniqueness result can be obtained 
as in Stewart \cite{Stewartuniq} by taking the difference of two solutions and showing 
that this difference is equal to zero by appliying Gronwall's inequality. The result 
we obtain thereby seems to be covered by the uniqueness theorem of Norris \cite{Norris}. 
Therefore, the proof by an independent method is of a minor interest and can be found 
in Cueto Camejo \cite{CuetoCamejo1}.

The paper is organized as follows. In Section $2$ we present the hypotheses for our 
problem and some necessary definitions. In Section $3$ we prove in Theorem 
\ref{existencetruncated} the existence and uniqueness of solutions to the truncated 
problem and we extract a weakly convergent subsequence in $L^1$ from a sequence of 
unique solutions for truncated equations to (\ref{problema})-(\ref{condinicial}). In 
Section $4$ we show that the solution of (\ref{problema}) is actually the limit 
function obtained from the weakly convergent subsequence of solutions of the truncated 
problem. In Section $5$ we prove the uniqueness, based on methods of Stewart 
\cite{Stewartuniq}, of the solutions to (\ref{problema})-(\ref{condinicial}) for a 
modification of the class (\ref{asterisco1}) of kernels. We obtain uniqueness for some 
kernels which are not covered by the existence result.
%
%
\section{Weak solutions in time and weighted \texorpdfstring{$L^1$}{L1}-spaces}
In order to study the existence of solutions of (\ref{problema})-(\ref{condinicial}), we 
define $Y$ to be the following Banach space with norm $\|\cdot\|_Y$
\begin{eqnarray*}
Y=\left\{u\in L^1\big(]0,\infty[\big): \|u\|_Y<\infty\right\}\quad\mbox{where}\quad\|u\|_Y=\int\limits_0^\infty(x+x^{-1})|u(x,t)|dx.
\end{eqnarray*}
That $Y$ is a Banach space is easily seen. We also write
\begin{eqnarray*}
\|u\|_x=\int\limits_0^\infty xu(x,t)dx\quad\mbox{and}\quad\|u\|_{x^{-1}}=\int\limits_0^\infty x^{-1}u(x,t)dx,
\end{eqnarray*}
and set
\begin{eqnarray*}
Y^+=\left\{u\in Y:u\geq 0 \quad a.e.\right\}.
\end{eqnarray*}
We define a solution of problem (\ref{problema})-(\ref{condinicial}) in the same 
way as Stewart ~\cite{Stewart}, i.e.\ solutions that are weak in time but classical 
in property space:
\begin{definition} Let $T\in]0,\infty[$. A solution $u(x,t)$ of (\ref{problema})-(\ref{condinicial}) is a function $u:[0,T[\longrightarrow Y^+$ such that for a.e. $x\in]0,\infty[$ and $t\in[0,T[$ the following properties hold
\begin{description}
\item[(i)]$u(x,t)\geq 0$ for all $t\in[0,\infty[$,
\item[(ii)]$u(x,\cdot)$ is continuous on $[0,T[$,
\item[(iii)]for all $t\in[0,T[$ the following integral is bounded
\begin{eqnarray*}
\int\limits_0^t\int\limits_0^\infty K(x,y)u(y,\tau)\,dy\,d\tau<\infty,
\end{eqnarray*}
\item[(iv)] for all $t\in[0,T[$, $u$ satisfies the following weak formulation of (\ref{problema})
\begin{eqnarray}
u(x,t)\!\!&\!\!=\!\!&\!\!u(x,0)+\int\limits_0^t\left[\frac{1}{2}\int\limits_0^xK(x-y,y)u(x-y,\tau)u(y,\tau)\,dy\right.\nonumber\\
&&\qquad\qquad\qquad\qquad\left.-\int\limits_0^\infty K(x,y)u(x,\tau)u(y,\tau)\,dy\right]d\tau.
\label{solutiondefinition}
\end{eqnarray}
\end{description}
\label{definicion}
\end{definition}
In the next sections we make use of the following hypotheses
\begin{hypotheses}
\begin{description}
  \item[]
  \item[(H1)]  $K(x,y)$ is a continuous non-negative function on $]0,\infty[\times]0,\infty[$, 
	\item[(H2)]  $K(x,y)$ is a symmetric function, i.e. $K(x,y)=K(y,x)$ for all $x,y\in]0,\infty[$, 
	\item[(H3)]  $K(x,y)\leq \kappa(1+x+y)^\lambda(xy)^{-\sigma}$ for $\sigma\in[0,1/2],\; \lambda-\sigma\in[0,1[$, and a constant $\kappa>0$.
\end{description}
\label{hypo}
\end{hypotheses}
In the rest of the paper we consider $\kappa=1$ for the simplicity.

We study the uniqueness of the solution to (\ref{problema})-(\ref{condinicial}) 
under the following further restriction on the kernels.

\begin{description}
\item[(H3')] $K(x,y)\leq\kappa_1(x^{-\sigma}+x^{\lambda-\sigma})(y^{-\sigma}+y^{\lambda-\sigma})$ \textit{such that} $\sigma,\;\lambda-\sigma\in[0,1/2]$ \textit{and} $\kappa_1>0$.
\end{description}

The restriction $\lambda-\sigma\in[0,1/2]$ in \textbf{(H3')} limits our uniqueness 
result to a subset of the kernels of the class defined in \textbf{(H3)}, namely to 
the ones for which $\lambda-\sigma\in[0,1/2]$ holds. But the class of kernels 
defined in \textbf{(H3')} is also wider than the defined in \textbf{(H3)} for 
$\lambda-\sigma\in[0,1/2]$. In this way we are also giving uniqueness result for 
kernels which are not included in the class defined in \textbf{(H3)}.

We introduce now some easily derived inequalities that will be used throughout the 
paper. The proof of these inequalities can be found in Giri \cite{AnkikThesis}. For 
any $x,y>0$
\begin{eqnarray}
2^{p-1}(x^p + y^p)\leq (x + y)^p \leq x^p + y^p\quad &\mbox{if} &\quad 0\leq p\leq 1, \label{inequality1}\\ 
2^{p-1}(x^p + y^p)\geq (x + y)^p\geq x^p + y^p\quad &\mbox{if} &\quad p\geq 1, \label{inequality2}\\ 
2^{p-1}(x^p + y^p)\geq (x + y)^p\quad &\mbox{if} &\quad p < 0. \label{inequality3} 
\end{eqnarray}
%
%
\section{The Truncated Problem}
$\quad$

We prove the existence of a solution to the problem (\ref{problema})-(\ref{condinicial}), 
by taking the limit of the sequence of solutions of the equations given by replacing 
the kernel $K(x,y)$ by the 'cut-off' kernel $K_n(x,y)$ for any given $n\in\mathbb{N}$,
\begin{equation*}
  K_n(x,y)=\left\{
                  \begin{array}{ll}
	                  K(x,y) & \mbox{if}\quad  x+y\leq n\quad\mbox{and}\quad x,y\geq 1/n \\ 
	                  0      & \mbox{otherwise}.
                  \end{array}
  \right.                
\end{equation*}
The resulting equations are written as
\begin{eqnarray}
  \dfrac{\partial u^n(x,t)}{\partial t}=\dfrac{1}{2}\int\limits_0^x{K_n(x-y,y)u^n(x-y,t)u^n(y,t)}\,dy - \int\limits_0^{n-x}{K_n(x,y)u^n(x,t)u^n(y,t)}\,dy, 
\label{cen} 
\end{eqnarray}
with the truncated initial data
\begin{eqnarray}
  u^n_0(x)=\left\{
                   \begin{array}{ll}
                     u_0(x) & \mbox{if}\quad 0\leq x\leq n \\ 
                     0      & \mbox{otherwise},
                   \end{array}
  \right. 
\label{icen1}                
\end{eqnarray}
where $u^n$ denotes the solution of the problem (\ref{cen})-(\ref{icen1}) for $x\in[0,n]$.
\begin{theorem}
  Suppose that \textbf{(H1)}, \textbf{(H2)}, \textbf{(H3)} hold and $u_0\in Y^+$. 
  Then for each $n=2,3,4,\ldots$ the problem (\ref{cen})-(\ref{icen1}) has a unique 
  solution $u^n$ with $u^n(x,t)\geq 0$ for a.e. $x\in[0,n]$ and $t\in [0,\infty[$. 
  Moreover, for all $t\in [0,\infty[$
  \begin{equation}
    \int_0^n{xu^n(x,t)}\,dx=\int_0^n{xu^n(x,0)}\,dx.
    \label{conservation}
  \end{equation}
  \label{existencetruncated}
\end{theorem}
The proof of Theorem \ref{existencetruncated} follows proceeding as in \cite[Theorem 3.1]{Stewart}
\subsection{Properties of the solutions of the truncated problem}

In the rest of the paper we consider for each $u^n$ their zero extension on $\mathbb R$, i.e.
\begin{eqnarray*}
\hat{u}^n(x,t)=\left\{
                      \begin{array}{ll}
                        u^n(x,t) & 0\leq x\leq n, \;\; t\in[0,T],\\
                        0        & x<0\;\;\mbox{or}\;\;x>n.
                      \end{array}
               \right.
\end{eqnarray*}
For clarity we drop the notation $\hat{\cdot}$ for the remainder of the paper.

\begin{lemma}
Assume that \textbf{(H1)}, \textbf{(H2)} and \textbf{(H3)} hold. Let us define 
$L=\|u_0^n\|_Y,$. We take $u^n$ to be the non-negative zero extension of the 
solution to the truncated problem found in Theorem \ref{existencetruncated}. 
Take any $T>0$. Then the following are true

\begin{description}
\item[(i)]We have uniformly for $t\in[0,T]$ the bound 
\begin{eqnarray*}\int\limits_0^\infty{(1+x+x^{-2\sigma})u^n(x,t)}\,dx \leq 3L.
\end{eqnarray*} 
\item[(ii)]Given $\epsilon>0$ there exists an $R>1$ such that for all $t\in[0,T]$
\begin{eqnarray*}
\sup_n\left\{\int\limits_R^\infty(1+x^{-\sigma})u^n(x,t)\,dx\right\}\leq\epsilon.
\end{eqnarray*}
\item[(iii)]Given $\epsilon>0$ there exists a $\delta>0$ such that for all $n=2,3,\ldots$ and $t\in\left[0,T\right]$
\begin{eqnarray*}
\int\limits_A (1+x^{-\sigma})u^n(x,t)\,dx<\epsilon \quad\quad\mbox{whenever}\quad\quad \mu(A)<\delta.
\end{eqnarray*}
\end{description}
\label{lemaproperty}
\end{lemma}
\textbf{Proof.}$\;$\textbf{Property (i)} We split the following integral into three parts
\begin{eqnarray}
\int\limits_0^\infty{(1+x+x^{-2\sigma})u^n(x,t)}\,dx=\int\limits_0^n{u^n(x,t)}\,dx + \int\limits_0^nxu^n(x,t)\,dx + \int\limits_0^n{x^{-2\sigma}u^n(x,t)}\,dx.
\label{unamas}
\end{eqnarray}
Working with the first integral of the right hand side of (\ref{unamas}) and using that $\sigma\in[0,\frac{1}{2}]$
\begin{eqnarray}
\int\limits_0^n{u^n(x,t)}\,dx \!\!&\!\!=\!\!&\!\! \int\limits_0^1{x^{-1}x u^n(x,t)}\,dx + \int\limits_1^n{x^{-1}x u^n(x,t)}\,dx \nonumber\\
                              \!\!&\!\!\leq\!\!&\!\! \int\limits_0^n{x^{-1} u^n(x,t)}\,dx + \int\limits_0^n{x u^n(x,t)}\,dx. \label{nonombre}
\end{eqnarray}
Now we proceed to obtain a uniform bound for the first term in the right hand 
side of (\ref{nonombre}). Multiplying equation (\ref{problema}) by $x^{-1}$ 
and integrating with respect to $x$ and $\tau$ from $0$ to $n$ and from $0$ 
to $t$ respectively, then changing the order of integration, then a change of 
variable $x-y=z$ and then re-changing the order of integration while replacing 
$z$ by $x$ gives 
\begin{eqnarray*}
\int\limits_0^n u^n(x,t)x^{-1}\,dx\!\!&\!\!=\!\!&\!\!\int\limits_0^t\left[\frac{1}{2}\int\limits_0^n\int\limits_0^x K_n(x-y,y)u^n(x-y,\tau)u^n(y,\tau)x^{-1}dy\,dx\right. \nonumber\\
\!\!&\!\!\!\!&\!\!\left. -\int\limits_0^n\int\limits_0^{n-x} K_n(x,y)u^n(x,\tau)u^n(y,\tau)x^{-1}dy\,dx\right]d\tau +\int\limits_0^n u_0^n(x)x^{-1}dx\nonumber \\
\!\!&\!\!=\!\!&\!\!\int\limits_0^t\left[\frac{1}{2}\int\limits_0^n\int\limits_0^{n-x}K_n(x,y)u^n(x,\tau)u^n(y,\tau)(x+y)^{-1}dy\,dx\right. \nonumber\\
\!\!&\!\!\!\!&\!\!\left. -\int\limits_0^n\int\limits_0^{n-x} K_n(x,y)u^n(x,\tau)u^n(y,\tau)x^{-1}dy\,dx\right]d\tau +\int\limits_0^n u_0^n(x)x^{-1}dx.
\end{eqnarray*}
Making use of the inequality (\ref{inequality3}) and the symmetry of $K(x,y)$ 
we obtain by omitting a negative term
\begin{eqnarray}
\int\limits_0^n u^n(x,t)x^{-1}dx
\!\!&\!\!\leq\!\!&\!\!\int\limits_0^t\left[\frac{1}{8}\int\limits_0^n\int\limits_0^{n-x}K_n(x,y)u^n(x,\tau)u^n(y,\tau)(x^{-1}+y^{-1})\,dy\,dx\right. \nonumber\\
\!\!&\!\!\!\!&\!\!\left. -\frac{1}{2}\int\limits_0^n\int\limits_0^{n-x} K_n(x,y)u^n(x,\tau)u^n(y,\tau)(x^{-1}+y^{-1})\,dy\,dx\right]d\tau +\int\limits_0^n u_0^n(x)x^{-1}dx\nonumber \\ 
\!\!&\!\!\leq\!\!&\!\!\int\limits_0^n u_0^n(x)x^{-1}dx\leq\|u_0^n\|_Y=L.
\label{un}
\end{eqnarray}
Using the mass conservation property (\ref{conservation}) and $n>1$ combined 
with (\ref{un}) brings (\ref{nonombre}) to
\begin{eqnarray}
\int\limits_0^n{u^n(x,t)}\,dx \leq\int\limits_0^n{x^{-1} u_0^n(x)}\,dx + \int\limits_0^n{x u_0^n(x)}\,dx\leq\|u_0^n\|_Y=L.
\label{47}
\end{eqnarray}
Now let us consider the third integral on the right hand side of (\ref{unamas})
\begin{eqnarray}
\int\limits_0^n{u^n(x,t)x^{-2\sigma}}\,dx \!\!&\!\!=\!\!&\!\! \int\limits_0^1u^n(x,t)x^{-2\sigma}\,dx+\int\limits_1^n u^n(x,t)x^{-2\sigma}\,dx \nonumber\\
                                       \!\!&\!\!\leq\!\!&\!\! \int\limits_0^nu^n(x,t)x^{-1}\,dx  + \int\limits_0^n xu^n(x,t)x\,dx \nonumber\\
                                       \!\!&\!\!\leq\!\!&\!\! \|u_0^n\|_Y=L.
\label{50}
\end{eqnarray}
Thus, by using (\ref{conservation}) together with (\ref{47}) and (\ref{50}) 
we may estimate
\begin{eqnarray*}
\int\limits_0^\infty (1+x+x^{-2\sigma})u^n(x,t)\,dx \leq 3\|u_0^n\|_Y=3L.
\end{eqnarray*}
\textbf{Property (ii)} Choose $\epsilon >0$ and let $R>1$ be such that 
$R>\frac{2\|u_0\|_Y}{\epsilon}$. Then we get using (\ref{conservation})
\begin{eqnarray*}
\int\limits_R^\infty(1+x^{-\sigma})u^n(x,t)\,dx = \int\limits_R^\infty(1+x^{-\sigma})\frac{x}{x}u^n(x,t)\,dx \!\!&\!\!\leq\!\!&\!\!\frac{1}{R}\int\limits_R^\infty(x+x^{1-\sigma})u^n(x,t)\,dx  \\
\!\!&\!\!\leq\!\!&\!\!\frac{2}{R}\|u^n_0\|_Y\leq\frac{2}{R}\|u_0\|_Y<\epsilon.
\end{eqnarray*}
\textbf{Property (iii)} By property \textbf{(ii)} we can choose $r>1$ such 
that for all $n$ and $t\in[0,T]$
\begin{eqnarray}
\int\limits_r^\infty (1+x^{-\sigma})u^n(x,t)\,dx<\frac{\epsilon}{2}.
\label{epsilon1}
\end{eqnarray}
Let $\chi_A$ denote the characteristic function of a set $A$, i.e.
\begin{eqnarray*}
\chi_A(x)=\left\{
\begin{array}{lcl}
1 & \mbox{if} & x\in A \\
0 & \mbox{if} & x\notin A.
\end{array}
\right.
\end{eqnarray*}
Let us define for all $n=1, 2, 3, \ldots$ and $t\in[0,T]$
\begin{eqnarray*}
f^n(A,t)=\sup_{0\leq z\leq r}\int\limits_0^\infty\chi_{A\cap[0,r]}(x+z)(1+x^{-\sigma})u^n(x,t)dx
\end{eqnarray*}
and set
\begin{eqnarray*}
k(r)=\frac{1}{2}\max_{\substack{0\leq x\leq r \\
                               0\leq y\leq r}
                     }(1+x+y)^{\lambda}(1+y^\sigma).
\end{eqnarray*}
Now, using $\|u_0^n\|_Y = L$ leads to 
\begin{eqnarray*}
\int\limits_0^\infty (1+x^{-\sigma})u_0^n(x)\,dx\leq2\int\limits_0^1 x^{-1}u_0^n(x)\,dx+2\int\limits_1^\infty xu_0^n(x)\,dx\leq 2\|u_0^n\|_Y =2L.
\end{eqnarray*}
By the absolute continuity of the Lebesgue integral, there exists a $\delta>0$ such that
\begin{eqnarray}
f^n(A,0)=\sup_{0\leq z\leq r}\int\limits_0^\infty\chi_{A\cap[0,r]}(x+z)(1+x^{-\sigma})u^n_0(x)\,dx < \frac{\epsilon}{2\exp\big(k(r)LT\big)},
\label{cond1}
\end{eqnarray}
whenever $A\subset]0,\infty[$ with $\mu(A)\leq\delta$. Now we multiply (\ref{cen}) by $\chi_{A\cap[0,r]}(x+z)(1+x^{-\sigma})$. This we integrate from $0$ to $t$ w.r.t. $s$ and over $[0,\infty[$ w.r.t. $x$. Using the non-negativity of each $u^n$ and $\mu(A)\leq\delta$ we obtain
\begin{eqnarray*}
\!\!&\!\!\!\!&\!\!\int\limits_0^\infty\chi_{A\cap[0,r]}(x+z)(1+x^{-\sigma})u^n(x,t)\,dx \\
\!\!&\!\!\!\!&\!\!\leq\frac{1}{2}\int\limits_0^t\int\limits_0^\infty\int\limits_0^\infty\chi_{A\cap[0,r]}(x+z)\chi_{[0,x]\cap[0,r]}(y)(1+x^{-\sigma})K_n(x-y,y)u^n(x-y,s)u^n(y,s)\,dy\,dx\,ds \\
\!\!&\!\!\!\!&\!\!\quad+\int\limits_0^\infty\chi_{A\cap[0,r]}(x+z)(1+x^{-\sigma})u^n_0(x)\,dx.
\end{eqnarray*}
Changing the order of integration, then making a change of variable $x-y=x'$ 
and replacing $x'$ by $x$ gives
\begin{eqnarray*}
\!\!&\!\!\!\!&\!\!\int\limits_0^\infty\chi_{A\cap[0,r]}(x+z)(1+x^{-\sigma})u^n(x,t)\,dx \\
\!\!&\!\!\!\!&\!\!\leq\frac{1}{2}\int\limits_0^t\int\limits_0^\infty\int\limits_0^\infty\chi_{A\cap[0,r]}(x+y+z)\chi_{[0,x+y]\cap[0,r]}(y)\left[1+(x+y)^{-\sigma}\right]K_n(x,y)u^n(x,s)u^n(y,s)\,dx\,dy\,ds \\
\!\!&\!\!\!\!&\!\!\quad+ \int\limits_0^\infty\chi_{A\cap[0,r]}(x+z)(1+x^{-\sigma})u^n_0(x)\,dx.
\end{eqnarray*}
Using the estimate \textbf{(H3)} of $K(x,y)$ we find
\begin{eqnarray*}
\!\!&\!\!\!\!&\!\!\int\limits_0^\infty\chi_{A\cap[0,r]}(x+z)(1+x^{-\sigma})u^n(x,t)\,dx \\
\!\!&\!\!\!\!&\!\!\leq\frac{1}{2}\int\limits_0^t\int\limits_0^\infty\int\limits_0^\infty\chi_{A\cap[0,r]}(x+y+z)\chi_{[0,x+y]\cap[0,r]}(y)\left[1+(x+y)^{-\sigma}\right](1+x+y)^{\lambda}(xy)^{-\sigma} \\
\!\!&\!\!\!\!&\!\!\qquad\qquad\qquad\qquad\qquad\qquad\qquad\qquad\qquad\qquad\qquad\qquad\cdot u^n(x,s)u^n(y,s)\,dx\,dy\,ds + f^n(A,0) \\
\!\!&\!\!\!\!&\!\!\leq\frac{1}{2}\int\limits_0^t\int\limits_0^\infty\int\limits_0^\infty\chi_{A\cap[0,r]}(x+y+z)\chi_{[0,x+y]\cap[0,r]}(y)(1+y^{-\sigma})(1+x+y)^{\lambda}(1+x^{-\sigma})y^{-\sigma} \\
\!\!&\!\!\!\!&\!\!\qquad\qquad\qquad\qquad\qquad\qquad\qquad\qquad\qquad\qquad\qquad\qquad\cdot u^n(x,s)u^n(y,s)\,dx\,dy\,ds + f^n(A,0) \\
\!\!&\!\!\!\!&\!\!=\frac{1}{2}\int\limits_0^t\int\limits_0^\infty\int\limits_0^\infty\chi_{A\cap[0,r]}(x+y+z)\chi_{[0,x+y]\cap[0,r]}(y)(1+y^{\sigma})(1+x+y)^{\lambda}(1+x^{-\sigma})y^{-2\sigma} \\
\!\!&\!\!\!\!&\!\!\qquad\qquad\qquad\qquad\qquad\qquad\qquad\qquad\qquad\qquad\qquad\qquad\cdot u^n(x,s)u^n(y,s)\,dx\,dy\,ds + f^n(A,0).
\end{eqnarray*}
We use now the definition of $k(r)$ and (\ref{50})
\begin{eqnarray*}
\!\!&\!\!\!\!&\!\!\int\limits_0^\infty\chi_{A\cap[0,r]}(x+z)(1+x^{-\sigma})u^n(x,t)\,dx \nonumber\\
\!\!&\!\!\!\!&\!\!\leq k(r)\int\limits_0^t\int\limits_0^ru^n(y,s)y^{-2\sigma}\int\limits_0^\infty\chi_{A\cap[0,r]}(x+y+z)(1+x^{-\sigma})u^n(x,s)\,dx\,dy\,ds + f^n(A,0) \nonumber\\
\!\!&\!\!\!\!&\!\!\leq k(r)\int\limits_0^t\int\limits_0^ru^n(y,s)y^{-2\sigma}\sup_{0\leq\omega\leq r}\int\limits_0^\infty\chi_{A\cap[0,r]}(x+\omega)(1+x^{-\sigma})u^n(x,s)\,dx\,dy\,ds + f^n(A,0) \nonumber\\
\!\!&\!\!\!\!&\!\!\leq k(r)L\int\limits_0^tf^n(A,s)\,ds + f^n(A,0).
\end{eqnarray*}
Since the right hand side is independent of $z$ we may take $\sup_{0\leq z\leq r}$ 
on the left hand side to obtain
\begin{eqnarray*}
f^n(A,t)\leq k(r)L\int\limits_0^tf^n(A,s)ds + \epsilon/\left(2\exp\big(k(r)LT\big)\right).
\end{eqnarray*}
By Gronwall's inequality, see e.g. Walter \cite[page 361]{Walter}
\begin{eqnarray}
f^n(A,t)\leq\epsilon\exp\big(k(r)LT\big)/\left(2\exp\big(k(r)LT\big)\right)=\frac{\epsilon}{2}.
\label{epsilon2}
\end{eqnarray}
By (\ref{epsilon1}) and (\ref{epsilon2}) follows that
\begin{eqnarray*}
\int\limits_A(1+x^{-\sigma})u^n(x,t)dx\!\!&\!\!=\!\!&\!\!\int\chi_{A\cap[0,r]}(x)(1+x^{-\sigma})u^n(x,t)dx + \int\chi_{A\cap[r,\infty[}(x)(1+x^{-\sigma})u^n(x,t)dx \\
                                      \!\!&\!\!\leq\!\!&\!\! f^n(A,t)+\int\limits_r^\infty(1+x^{-\sigma})u^n(x,t)dx \\
                                      \!\!&\!\!\leq\!\!&\!\! \epsilon/2 + \epsilon/2 = \epsilon 
\end{eqnarray*}
whenever $\mu(A)<\delta$. 

This completes the proof of Lemma \ref{lemaproperty}. \hfill{$\Box$}

Let us define $v^n(x,t)=x^{-\sigma}u^n(x,t)$. Due to the Lemma \ref{lemaproperty} 
above and the Dunford-Pettis Theorem \cite[page 274]{Edwards}, we can conclude that 
for each $t\in[0,T]$ the sequences $\big(u^n(t)\big)_{n\in\mathbb N}$ and 
$\big(v^n(t)\big)_{n\in\mathbb N}$ are weakly relatively compact in $L^1\big(]0,\infty[\big)$.
%
%
\subsection{Equicontinuity in time}

\begin{lemma}
Assume that Hypotheses $1.1$ hold. Take $\big(u^n\big)$ now to be the sequence of extended 
solutions to the truncated problems (\ref{cen})-(\ref{icen1}) found in Theorem 
\ref{existencetruncated} and $v^n(x,t)=x^{-\sigma}u^n(x,t)$. Then there exists a subsequences 
$\big(u^{n_k}(t)\big)$ and  $\big(v^{n_l}(t)\big)$ of $\big(u^{n}(t)\big)_{n\in\mathbb N}$ and 
$\big(v^{n}(t)\big)_{n\in\mathbb N}$ respectively such that
\begin{eqnarray*}
&& u^{n_k}(t)\rightharpoonup u(t)\quad\mbox{in}\quad L^1\big(]0,\infty[\big)\quad\mbox{as}\quad n_k\rightarrow\infty \\ 
&& v^{n_l}(t)\rightharpoonup v(t)\quad\mbox{in}\quad L^1\big(]0,\infty[\big)\quad\mbox{as}\quad n_l\rightarrow\infty 
\end{eqnarray*}
for any $t\in [0,T]$. This convergence is uniform for all $t\in [0,T]$ giving 
$u,v\in C_B\left([0,\infty[;\Omega\right)=\left\{\eta:[0,\infty[\rightarrow \Omega,\right.$ $\left.\eta \;\mbox{continuous and}\;\eta(t)\;\mbox{bounded for all}\;t\geq 0\right\}$, and where $\Omega$ is $L^1\left(]0,\infty[\right)$ \\equipped with the weak topology. \label{unlema}
\end{lemma}
\textbf{Proof}: Choose $\epsilon>0$ and $\phi\in L^\infty\big(]0,\infty[\big)$. Let $s,t\in[0,T]$ and assume that $t\geq s$. Choose $a>1$ such that 
\begin{eqnarray}
\frac{6L}{a}\|\phi\|_{L^\infty\left(]0,\infty[\right)}\leq\epsilon/2.
\label{20}
\end{eqnarray}
Let us define the function $\omega^n(x,t):=u^n(x,t)x^{-\beta}$ for $\beta=0$ and $\beta=\sigma$. Then we have that for $\beta=0$ and $\beta=\sigma$ the function $\omega$ becames $u^n(x,t)$ and $v^n(x,t)$ respectively. 
Using Lemma \ref{lemaproperty}, 
for each $n$, we get using $a>1$ chosen to satisfy (\ref{20})
\begin{eqnarray}
\int\limits_a^\infty\left|\omega^n(x,t)-\omega^n(x,s)\right|dx\!\!&\!\!=\!\!&\!\!\int\limits_a^\infty\left|x^{-\beta}u^n(x,t)-x^{-\beta}u^n(x,s)\right|dx \nonumber\\
                                                  \!\!&\!\!\leq\!\!&\!\!\frac{1}{a}\int\limits_a^\infty x^{1-\beta}\left|u^n(x,t)+u^n(x,s)\right|dx \nonumber\\
                                                  \!\!&\!\!\leq\!\!&\!\!\frac{1}{a}\int\limits_a^\infty x\left|u^n(x,t)+u^n(x,s)\right|dx \leq 6L/a.
\label{otratu}
\end{eqnarray}
By using (\ref{cen}), (\ref{20}), (\ref{otratu}), for $t\geq s$ and the definition of $\omega^n(x)$ we obtain
\begin{eqnarray*}
\!\!&\!\!\!\!&\!\!\left|\int\limits_0^\infty\phi(x)\left[\omega^n(x,t)-\omega^n(x,s)\right]dx\right|\\
\!\!&\!\!\!\!&\!\!\leq\int\limits_0^a\left|\phi(x)\right|\left|\omega^n(x,t)-\omega^n(x,s)\right|dx+ \epsilon/2 \\ \!\!&\!\!\!\!&\!\!\leq\|\phi\|_{L^\infty\left(]0,\infty[\right)}\int\limits_s^t\left[\frac{1}{2}\int\limits_0^a\int\limits_0^xK_n(x-y,y)u^n(x-y,\tau)u^n(y,\tau)x^{-\beta}dy\,dx\right. \\
\!\!&\!\!\!\!&\!\!\qquad\qquad\qquad\qquad\quad\left.+\int\limits_0^a\int\limits_0^{n-x}K_n(x,y)u^n(x,\tau)u^n(y,\tau)x^{-\beta}dy\,dx\right]d\tau + \epsilon/2 \\
\!\!&\!\!\!\!&\!\!=\|\phi\|_{L^\infty\left(]0,\infty[\right)}\int\limits_s^t\left[\frac{1}{2}\int\limits_0^a\int\limits_0^{a-x}K_n(x,y)u^n(x,\tau)u^n(y,\tau)(x+y)^{-\beta}dy\,dx\right. \\ \!\!&\!\!\!\!&\!\!\qquad\qquad\qquad\qquad\quad\left.+\int\limits_0^a\int\limits_0^{n-x}K_n(x,y)u^n(x,\tau)u^n(y,\tau)x^{-\beta}dy\,dx\right]d\tau + \epsilon/2.
\end{eqnarray*}
Taking $y=0$ in the term $(x+y)^{-\beta}$ we proceed as follows
\begin{eqnarray*}
\!\!&\!\!\!\!&\!\! \left|\int\limits_0^\infty\phi(x)\left[\omega^n(x,t)-\omega^n(x,s)\right]dx\right| \nonumber\\
\!\!&\!\!\!\!&\!\!\leq\|\phi\|_{L^\infty\left(]0,\infty[\right)}\int\limits_s^t\left[\frac{1}{2}\int\limits_0^a\int\limits_0^{a-x}K_n(x,y)u^n(x,\tau)u^n(y,\tau)x^{-\beta}dy\,dx\right.\\
\!\!&\!\!\!\!&\!\!\qquad\qquad\qquad\qquad\quad\left. + \int\limits_0^a\int\limits_0^{n-x}K_n(x,y)u^n(x,\tau)u^n(y,\tau)x^{-\beta}dy\,dx\right]d\tau + \epsilon/2 \nonumber\\                     
\!\!&\!\!\!\!&\!\!\leq\frac{3}{2}\|\phi\|_{L^\infty\left(]0,\infty[\right)}\int\limits_s^t\int\limits_0^\infty\int\limits_0^\infty K(x,y)u^n(x,\tau)u^n(y,\tau)x^{-\beta}dy\,dx + \epsilon/2. 
\label{arriba5} 
\end{eqnarray*}
Now we use of the inequalities (\ref{inequality1}) and (\ref{inequality2}) to obtain the following
\begin{eqnarray}
(1+x+y)^p\leq C(1+x^p+y^p)\quad\mbox{where}\quad C=\left\{
                                                          \begin{array}{lcl}
                                                           1 & \mbox{if} & 0\leq p\leq 1 \\
                                                           2^{2p-2} & \mbox{if} & p>1,
                                                          \end{array}
                                                   \right.
\label{C}
\end{eqnarray}
Using the estimation of $K(x,y)$ and the inequality (\ref{inequality1}) together (\ref{C}) for $p=\lambda$ we have
\begin{eqnarray*}
\!\!&\!\!\!\!&\!\! \left|\int\limits_0^\infty\phi(x)\left[\omega^n(x,t)-\omega^n(x,s)\right]dx\right|  \\ 
\!\!&\!\!\!\!&\!\!\leq\frac{3}{2}\|\phi\|_{L^\infty\left(]0,\infty[\right)}\int\limits_s^t\int\limits_{0}^\infty\int\limits_{0}^\infty(1+x+y)^\lambda(xy)^{-\sigma} u^n(x,\tau)u^n(y,\tau)x^{-\beta}dy\,dx\,d\tau +\epsilon/2\\
\!\!&\!\!\!\!&\!\!\leq \frac{3}{2}C\|\phi\|_{L^\infty\left(]0,\infty[\right)}\int\limits_s^t\int\limits_{0}^\infty\int\limits_{0}^\infty(x^{-(\sigma+\beta)}y^{-\sigma}+x^{\lambda-(\sigma+\beta)}y^{-\sigma}+y^{\lambda-\sigma}x^{-(\sigma+\beta)})u^n(x,\tau)u^n(y,\tau)\,dy\,dx\,d\tau\\&&\qquad+\epsilon/2.
\end{eqnarray*}
By using Lemma \ref{lemaproperty}(\textit{i}) we obtain
\begin{eqnarray} 
\left|\int\limits_0^\infty\phi(x)\left[\omega^n(x,t)-\omega^n(x,s)\right]dx\right| \leq\frac{27}{2}C\|\phi\|_{L^\infty\left(]0,\infty[\right)}(t-s)L^2+\epsilon/2<\epsilon
\label{pag15}
\end{eqnarray}
whenever $(t-s)<\delta$ for some $\delta>0$ sufficiently small. The argument given above 
similarly holds for $s<t$. Hence (\ref{pag15}) holds for all $n$ and $|t-s|<\delta$. Then the 
sequence $\big(\omega^n(t)\big)_{n\in\mathbb N}$ is time equicontinuous in $L^1\big(]0,\infty[\big)$. 
Thus, $\big(\omega^n(t)\big)$ lies in a relatively compact subset of a gauge space $\Omega_1$. 
The gauge space $\Omega_1$ is $L^1\big(]0,\infty[\big)$ equipped with the weak topology. For details 
about gauge spaces, see Ash ~\cite[page 226]{Ash}. Then, we may apply a version of the 
Arzela-Ascoli Theorem, see Ash ~\cite[page 228]{Ash}, to conclude that there exists a subsequence 
$\big(\omega^{n_k}\big)_{k\in\mathbb N}$ such that
\begin{eqnarray*}
\omega^{n_k}(t)\rightarrow \omega(t)\quad\mbox{in}\quad \Omega\quad\mbox{as}\quad n_k\rightarrow\infty,
\end{eqnarray*}
uniformly for $t\in [0,T]$ for some $\omega\in C\left([0,T];\Omega\right)$.
Then taking $\beta=0$ and $\beta=\sigma$ we can conclude that there exist subsequences $\big(u^{n_k}\big)_{k\in\mathbb N}$ and $\big(v^{n_k}\big)_{k\in\mathbb N}$ such that 
\begin{eqnarray*}
u^{n_k}(t)\rightarrow u(t)\quad\mbox{in}\quad \Omega\quad\mbox{as}\quad n_k\rightarrow\infty,\\
v^{n_k}(t)\rightarrow v(t)\quad\mbox{in}\quad \Omega\quad\mbox{as}\quad n_k\rightarrow\infty,
\end{eqnarray*}
uniformly for $t\in [0,T]$ for some $u,v\in C\left([0,T];\Omega\right)$.\\
Since $T>0$ is arbitrary we obtain $u,v\in C_B\left([0,\infty[;\Omega\right)$\hfill{$\Box$}
\begin{lemma}
For $v^n(\cdot,t)$ defined as before, we have
\begin{eqnarray*}
v^n(\cdot,t)\rightharpoonup v(\cdot,t)\quad\mbox{where}\quad v(x,t)=x^{-\sigma}u(x,t)\quad \mbox{for all}\quad t\in[0,T]\quad\mbox{in}\quad L^1(]0,a]).
\end{eqnarray*}
\end{lemma}
\textbf{Proof.} By Lemma \ref{unlema}, we know that $v^{n}(t)\rightharpoonup v(t)\quad\mbox{in}\quad L^1\big(]0,\infty[\big)\quad\mbox{as}\quad n\rightarrow\infty$ uniformly for $t\in [0,T]$. Then, we just need to prove that $v(x,t)=x^{-\sigma}u(x,t)$

By definition of weak convergence we have
\begin{eqnarray*}
\int\limits_0^a\varphi(x)\left[v^n(x,t)-v(x,t)\right]dx\rightarrow 0\quad\mbox{for all}\quad \varphi\in L^\infty\big(]0,a]\big)
\end{eqnarray*}
as $x^{\sigma}\in L^\infty\big(]0,a]\big)$
\begin{eqnarray*}
\int\limits_0^a\varphi(x)\left[x^{\sigma}v^n(x,t)-x^{\sigma}v(x,t)\right]dx=\int\limits_0^a\varphi(x)\left[u^n(x,t)-x^{\sigma}v(x,t)\right]dx\rightarrow 0
\end{eqnarray*}
for all $\varphi\in L^\infty\big(]0,a]\big)$ as $u^n\rightharpoonup u$ we have due to the uniqueness of the weak limit of weak convergence, $v(x,t)=x^{-\sigma}u(x,t)$.\hfill{$\Box$}
%
%
\section{Existence Theorem}
\subsection{Convergence of the integrals}

In order to show that the limit function which we obtained above is indeed a solution to (\ref{problema})-(\ref{condinicial}), we define the
operators $M^n_i$, $M_i$, $i=1,2$
\begin{eqnarray*}
M_1^n(u^n)(x)\!\!&\!\!=\!\!&\!\!\frac{1}{2}\int\limits_{0}^xK_n(x-y,y)u^n(x-y)u^n(y)\,dy \\M_1(u)(x)&=&\frac{1}{2}\int\limits_0^xK(x-y,y)u(x-y)u(y)\,dy \\ 
M_2^n(u^n)(x)\!\!&\!\!=\!\!&\!\!\int\limits_{0}^{n-x}K_n(x,y)u^n(x)u^n(y)\,dy \\ M_2(u)(x)&=&\int\limits_0^\infty K(x,y)u(x)u(y)\,dy,
\end{eqnarray*}
where $u\in L^1\big(]0,\infty[\big)$, $x\in[0,\infty[$ and $n=1, 2, \ldots$. Set $M^n=M^n_1-M^n_2$ and $M=M_1-M_2$.
\begin{lemma}
Suppose that $\big(u^n\big)_{n\in\mathbb N}\subset Y^+$, $u\in Y^+$ where $\|u^n\|_Y\leq L$, 
$\|u\|_Y\leq Q$, $u^n\rightharpoonup u$ and $v^n\rightharpoonup v$ in $L^1\big(]0,\infty[\big)$ as 
$n\rightarrow\infty$. Then for each $a>0$
\begin{eqnarray*}
M^n(u^n)\rightharpoonup M(u)\quad\mbox{in}\quad L^1\big(]0,a[\big)\quad\mbox{as}\quad n\rightarrow\infty. 
\end{eqnarray*}
\label{lemaconvergence}
\end{lemma}
\textbf{Proof}: Choose $a>0$ and let $\phi\in L^\infty\big(]0,\infty[\big)$. We show that 
$M^n_i(u^n)\rightharpoonup M_i(u)$ in $L^1\big(]0,a[\big)$ as $n\rightarrow\infty$ for $i=1,2.$ \newline
The proof of case $\textit{i}=1$ is analogous to the proof of the $W_1$ case in \cite[Lemma 4.1]{Stewart} by taking
\begin{eqnarray*}
g(v)(x)=\frac{1}{2}\int\limits_0^{a-x}\phi(x+y)K(x,y)(xy)^\sigma v(y)\,dy\quad\mbox{where}\quad v=x^{-\sigma}u. 
\end{eqnarray*}
 For every $\epsilon>0$ and $C$ defined by (\ref{C}) we can choose $b$ such that
\begin{eqnarray}
C\|\phi\|_{L^\infty\left([0,a]\right)}\left[(2b^{-(1+\sigma)}+b^{\lambda-\sigma-1})(L^2+Q^2)\right]<\frac{\epsilon}{3}
\label{asterisco3prima}
\end{eqnarray}
Redefining the operator $g$ for $u\in Y^+$ and $x\in[0,a]$ by
\begin{eqnarray*}
g(v)(x)=\int\limits_0^b\phi(x)K(x,y)(xy)^\sigma v(y)\,dy.
\end{eqnarray*}
We can now follow the lines of the proof of the $W_2$ case in \cite[Lemma 4.1]{Stewart} to get the proof of case $\textit{i}=2$.

Then the proof of Lemma \ref{lemaconvergence} is complete.
%
%
\subsection{The existence result}

\begin{theorem}
Suppose that Hypotheses \ref{hypo} hold and assume that $u_0\in Y^+$. Then (\ref{solutiondefinition}) has a 
solution $u\in C_B\left([0,\infty[,L^1\big(]0,\infty[\big)\right)$. Moreover, we also obtain $u\in C_B^1\left([0,\infty[,L^1\big(]0,\infty[\big)\right)$
and therefore $u$ is a \emph{regular} solution satisfying (\ref{problema})
\label{existen}
\end{theorem}
\textbf{Proof.} Choose $T,m > 0$, and let $\big(u^n\big)_{n\in\mathbb N}$ be the weakly convergent 
subsequence of approximating solutions obtained above, in the proof of Lemma \ref{unlema}. From 
Lemma \ref{unlema} we have $u\in C_B\left([0,\infty[,\Omega\right)$. For 
$t\in[0,T]$ we obtain due to weak convergence
\begin{eqnarray*}
\int\limits_0^mxu(x,t)\,dx=\lim_{n\rightarrow\infty}\int\limits_0^mxu^n(x,t)\,dx\quad\mbox{and}\quad\int\limits_{1/m}^mx^{-1}u(x,t)\,dx=\lim_{n\rightarrow\infty}\int\limits_{1/m}^mx^{-1}u^n(x,t)\,dx.
\end{eqnarray*}
Using the mass conservation property (\ref{conservation}) and (\ref{un}), this gives the uniform estimate
\begin{eqnarray*}
\int\limits_0^mxu(x,t)\,dx+\int\limits_{1/m}^mx^{-1}u(x,t)\,dx\leq 2L\quad\mbox{for any}\; n\in\mathbb{N}.
\end{eqnarray*}
Then taking $m\rightarrow\infty$ the uniqueness of weak limits implies that $u\in Y^+$ with 
$\|u\|_Y\leq 2L$. Let $\phi\in L^\infty\big(]0,a[\big)$. From Lemma \ref{unlema} we have for each 
$s\in[0,t]$
\begin{eqnarray}
u^n(t)\rightharpoonup u(t)\quad\mbox{in}\quad L^1(]0,a[)\quad\mbox{as}\quad n\rightarrow\infty.
\label{3232}
\end{eqnarray}
For Lemma \ref{unlema} and Lemma \ref{lemaconvergence} for each $s\in[0,t]$ we have for $M^n=M_1^n-M_2^n$ and $M=M_1-M_2$
\begin{eqnarray}
\int\limits_0^a\phi(x)\left[M^n(u^n(s))(x)-M(u(s))(x)\right]dx\rightarrow 0\quad\mbox{as}\quad n\rightarrow\infty.
\label{341}
\end{eqnarray}
Also, for $s\in[0,t]$, using Lemma \ref{lemaproperty}\textbf{(i)}, $\|u\|_Y\leq 2L$, and $C$ as in (\ref{C}) we find that
\begin{eqnarray}
&&\int\limits_0^a\left|\phi(x)\right|\left|M^n(u^n(s))(x)-M(u(s))(x)\right|dx \nonumber\\
&&\leq\|\phi\|_{L^\infty\left(]0,a[\right)}\left[\frac{1}{2}\int\limits_0^a\int\limits_0^xK(x-y,y)\left[u^n(x-y,s)u^n(y,s)+u(x-y,s)u(y,s)\right]dy\,dx\right. \nonumber\\ 
&&\qquad\qquad\qquad\quad+\left.\int\limits_0^a\int\limits_0^{n-x}K(x,y)u^n(x,s)u^n(y,s)\,dy\,dx + \int\limits_0^a\int\limits_0^\infty K(x,y)u(x,s)u(y,s)\,dy\,dx\right] \nonumber\\
&&\leq\|\phi\|_{L^\infty\left(]0,a[\right)}\left[\frac{5}{2}(1+2a)^\lambda+19C\right]L^2.
\label{351}
\end{eqnarray}
Since the left-hand side of (\ref{351}) is in $L^1\big(]0,t[\big)$ we have by (\ref{341}), (\ref{351}) and the dominated convergence theorem
\begin{eqnarray}
\left|\int\limits_0^t\int\limits_0^a\phi(x)\left[M^n(u^n(s))(x)-M(u(s))(x)\right]dx\,ds\right|\rightarrow 0\quad\mbox{as}\quad n\rightarrow\infty.
\label{361}
\end{eqnarray}
Since $\phi$ was chosen arbitrarily the limit (\ref{361}) holds for all $\phi\in L^\infty\big(]0,a[\big)$. By Fubini's Theorem we get
\begin{eqnarray}
\int\limits_0^t M^n(u^n(s))(x)\,ds\rightharpoonup\int\limits_0^tM(u(s))(x)\,ds\quad\mbox{in}\quad L^1\big(]0,a[\big)\quad\mbox{as}\quad n\rightarrow\infty.
\label{371}
\end{eqnarray}
From the definition of $M^n$ for $t\in[0,T]$
\begin{eqnarray*}
u^n(t)=\int\limits_0^t M^n(u^n(s))\,ds + u^n(0).
\end{eqnarray*}
Thus it follows by (\ref{371}), (\ref{3232}) and the uniqueness of weak limits that for all $t\in [0,T]$
\begin{eqnarray}
u(x,t)=\int\limits_0^t M(u(s))(x)\,ds + u(x,0)\quad\mbox{for a.e.}\quad x\in[0,a]. 
\label{hayqueusarte}
\end{eqnarray}
It follows from the fact that $T$ and $a$ are arbitrary that $u$ is a solution to (\ref{problema}) 
in $u\in C_B\big([0,\infty[,\Omega\big)$. 

In order to show that $u\in C_B\left([0,\infty[,L^1\big(]0,\infty[\big)\right)$ we consider w.l.o.g. $t_n > t$ and by using (\ref{hayqueusarte}) we have that
\begin{eqnarray*}
\int\limits_0^\infty|u(x,t_n)-u(x,t)|dx\!\!&\!\!=\!\!&\!\!\int\limits_0^\infty\left|\frac{1}{2}\int\limits_t^{t_n}\int\limits_0^xK(x-y,y)u(x-y,\tau)u(y,\tau)dy\,d\tau\right.\nonumber\\
&&\qquad\qquad\left.-\int\limits_t^{t_n}\int\limits_0^\infty K(x,y)u(x,\tau)u(y,\tau)dy\,d\tau\right|dx\nonumber\\
&\leq&\frac{3}{2}\int\limits_t^{t_n}\int\limits_0^\infty\int\limits_0^\infty K(x,y)u(x,\tau)u(y,\tau)dy\,dx\,d\tau\nonumber\\
\end{eqnarray*}
By using the definition (\ref{C}) of $C$ and $\|u\|_Y\leq 2L$ we find that
\begin{eqnarray}
\int\limits_0^\infty|u(x,t_n)-u(x,t)|dx&\leq&\frac{3}{2}\int\limits_t^{t_n}\int\limits_0^\infty\int\limits_0^\infty  (1+x+y)^\lambda(xy)^{-\sigma}u(x,\tau)u(y,\tau)\,dy\,dx\nonumber\\
\!\!&\!\!\leq\!\!&\!\!\frac{3}{2}C\int\limits_t^{t_n}\int\limits_0^\infty\int\limits_0^\infty(1+x^\lambda+y^\lambda)(xy)^{-\sigma}u(x,\tau)u(y,\tau)\,dy\,dx\nonumber\\
\!\!&\!\!=\!\!&\!\!\frac{3}{2}C\int\limits_t^{t_n}\int\limits_0^\infty\int\limits_0^\infty\left[(xy)^{-\sigma}+x^{\lambda-\sigma}y^{-\sigma}+y^{\lambda-\sigma} x^{-\sigma}\right]u(x,\tau)u(y,\tau)\,dy\,dx\nonumber\\
&\leq&18CL^2(t_n-t).
\label{L1}
\end{eqnarray}
Then from (\ref{L1}) we obtain that
\begin{eqnarray}
\int\limits_0^\infty|u(x,t_n)-u(x,t)|dx\rightarrow 0\quad\mbox{as}\quad t_n\rightarrow t.
\label{continuidad}
\end{eqnarray}
The same argument holds when $t_n<t$. Hence (\ref{continuidad}) holds for $|t_n-t|\rightarrow 0$ and we can conclude that $u\in C_B\left([0,\infty[,L^1\big(]0,\infty[\big)\right)$.

Now, we have that our solution satisfies the integral equation
\begin{eqnarray}
u(x,t)\!\!&\!\!=\!\!&\!\!u(x,0)+\int\limits_0^t\left[\frac{1}{2}\int\limits_0^xK(x-y,y)u(x-y,\tau)u(y,\tau)\,dy\right.\nonumber\\
&&\qquad\qquad\qquad\qquad\left.-\int\limits_0^\infty K(x,y)u(x,\tau)u(y,\tau)\,dy\right]d\tau.
\label{asteriscounavezmas}
\end{eqnarray}
From this we can see that for $u$, which is a continuous function in time $t$, that the integrand
\begin{eqnarray}
f(x,t)=\frac{1}{2}\int\limits_0^xK(x-y,y)u(x-y,t)u(y,t)\,dy-\int\limits_0^\infty K(x,y)u(x,t)u(y,t)\,dy
\label{derivadaL1}
\end{eqnarray}
is also a continuous function in time. We now show that $f(\cdot,t)\in L^1\big([0,\infty[\big)$ for any $t\in[0,\infty[$.

Integrating (\ref{derivadaL1}) from $0$ to $\infty$ w.r.t.\ $x$ we have to show that the following
integral is bounded
\begin{eqnarray}
 \int\limits_0^\infty f(x,t)\,dx\!\!&\!\!=\!\!&\!\!\frac{1}{2}\int\limits_0^\infty\int\limits_0^xK(x-y,y)u(x-y,t)u(y,t)\,dy\,dx\nonumber\\
 &&\qquad\qquad\qquad-\int\limits_0^\infty\int\limits_0^\infty K(x,y)u(x,t)u(y,t)\,dy\,dx.
\label{integralL1}
\end{eqnarray}
Working with the second term of the right hand side of (\ref{integralL1}) as in (\ref{L1}) we find that 
\begin{eqnarray}
\int\limits_0^\infty\int\limits_0^\infty K(x,y)u(x,\tau)u(y,\tau)\,dy\,dx<\infty.
\label{L11}
\end{eqnarray}
Now, by Tonelli's Theorem \cite[page 293]{Nielsen} we have that 
\begin{eqnarray*}
 \int\limits_0^\infty\int\limits_0^xK(x-y,y)u(x-y,\tau)u(y,\tau)\,dy\,dx=\int\limits_0^\infty\int\limits_y^\infty K(x-y,y)u(x-y,\tau)u(y,\tau)\,dx\,dy
\end{eqnarray*}
holds if
\begin{eqnarray*}
 \int\limits_0^\infty\int\limits_0^xK(x-y,y)u(x-y,\tau)u(y,\tau)\,dy\,dx<\infty
\end{eqnarray*}
or
\begin{eqnarray*}
\int\limits_0^\infty\int\limits_y^\infty K(x-y,y)u(x-y,\tau)u(y,\tau)\,dx\,dy<\infty.
\end{eqnarray*}
Making a change of varible $x-y=x'$, $y=y'$ in the second integral term we find, by using the 
symmetry of $K(x,y)$, that using (\ref{L11})
\begin{eqnarray*}
 \quad\int\limits_0^\infty\int\limits_y^\infty K(x-y,y)u(x-y,\tau)u(y,\tau)\,dx\,dy\!\!&\!\!=\!\!&\!\!\int\limits_0^\infty\int\limits_0^\infty K(x',y')u(x',\tau)u(y',\tau)\,dx'\,dy'\\
 \!\!&\!\!=\!\!&\!\!\int\limits_0^\infty\int\limits_0^\infty K(x',y')u(x',\tau)u(y',\tau)\,dy'\,dx'<\infty.
\end{eqnarray*}
From this it follows that
\begin{eqnarray}
 \int\limits_0^\infty\int\limits_0^xK(x-y,y)u(x-y,\tau)u(y,\tau)\,dy\,dx=\int\limits_0^\infty\int\limits_0^\infty K(x',y')u(x',\tau)u(y',\tau)\,dy'\,dx'<\infty.
 \label{L12}
\end{eqnarray}
Then, from (\ref{L11}) and (\ref{L12}) together with (\ref{integralL1}) it follows that 
$f(\cdot,t)\in L^1\big([0,\infty[\big)$. Moreover, we have that $f\in C_B\left([0,\infty[,L^1\big(]0,\infty[\big)\right)$. Then, using this fact, (\ref{asteriscounavezmas}), (\ref{derivadaL1}), and $u(x,0)\in Y^+$ we find that
\begin{eqnarray*}
 u(x,t)\!\!&\!\!=\!\!&\!\!u(x,0)+\int\limits_0^t\left[\frac{1}{2}\int\limits_0^xK(x-y,y)u(x-y,\tau)u(y,\tau)\,dy\right.\\
 &&\qquad\qquad\qquad\qquad\left.-\int\limits_0^\infty K(x,y)u(x,\tau)u(y,\tau)\,dy\right]d\tau,
\end{eqnarray*}
gives $u\in C_B^1\left([0,\infty[,L^1\big(]0,\infty[\big)\right)$ since the right hand side lies in this space. And this completes the proof of Theorem \ref{existen}.\hfill{$\Box$}
%
%
\section{Uniqueness of Solutions}
\begin{theorem}
If \textbf{(H1)}, \textbf{(H2)} and \textbf{(H3')} hold then the problem 
(\ref{problema})-(\ref{condinicial}) has a unique solution $u\in C_B\left([0,\infty[,L^1\big(]0,\infty[\big)\right)$.
\end{theorem}
This result seems to be covered by the uniqueness theorem of Norris \cite{Norris}. Therefore the proof by an independent method is of minor interest and can be found in Cueto Camejo \cite{CuetoCamejo1}.
%
%
\phantomsection
\section*{Acknowledgements}

This work was supported by the International Max-Planck Research School, `Analysis, Design and Optimization in Chemical and Biochemical Process
Engineering', Otto-von-Guericke-Universit\"at Magdeburg. The authors gratefully thank for the funding of C. Cueto Camejo through this PhD program by the state of Saxony-Anhalt. 

\end{document}